\documentclass[11pt]{article}
\usepackage{amssymb,amsmath,graphicx,color}
\usepackage[top=1in, bottom=1in, left=1in, right=1in]{geometry}
\usepackage{graphicx,float,amsmath}

\graphicspath{ {./pics/} }

\begin{document}

\begin{titlepage}
\clearpage\thispagestyle{empty}

\begin{centering}
\large{\textbf{Reynolds number scaling of influence of boundary layers on
the \\ global behavior of laboratory quasi-Keplerian flows}}\\

\vspace{1cm}

Eric M. Edlund* and Hantao Ji \\
Princeton Plasma Physics Laboratory \\

\vspace{0.5cm}

September 1, 2015

\vspace{1cm}

\end{centering}

\begin{abstract}
We present measurements of quasi-Keplerian flows in a Taylor-Couette device
that identify the boundary conditions required to generate near-ideal flows
that exhibit self-similarity under scaling of the Reynolds number. These
experiments are contrasted with alternate boundary configurations that result
in flows that progressively deviate from ideal Couette rotation as the
Reynolds number is increased. These behaviors are quantitatively explained
in terms of the tendency to generate global Ekman circulation and the
balance of angular momentum fluxes through the axial and radial boundary
layers.
\end{abstract}

\vspace{2 cm}

\hspace{0.5cm}{*current email address: eedlund@mit.edu}

\end{titlepage}

\clearpage

\section{Introduction}

That the global properties of an extended system may be mapped to the
boundaries is an idea that has found success in holographic theories of general
relativistic systems \cite{Adams,Carrasco}, and in magnetically confined
plasmas \cite{Rice,Gurcan}. We report on a similar behavior observed in
incompressible hydrodynamic flows in a Taylor-Couette apparatus where it is
observed that certain characteristics of the global flow are largely dictated by
the boundaries. This finding is particularly relevant for experiments that
examine quasi-Keplerian (QK) flows, that is, rotation satisfying $0 < q < 2$
where $q = - d \ln \Omega/d \ln r$, $\Omega$ is the fluid angular velocity and
$r$ is the radial coordinate, as models of astrophysical systems, namely
accretion disks. Numerous recent studies have commented on the
hydrodynamic stability of such systems \cite{Ji_Nature, SchartmanAA,
Edlund_PRE, Paoletti}, with extensions to magnetohydrodynamics in
electrically conducting fluids \cite{Seilmeyer, Stefani, Sisan, Roach, Spence}.

While there is some disagreement between studies as to whether
hydrodynamic turbulence can be induced in QK flows, the balance seems to
lean toward the negative, at least insofar as incompressible turbulence is
considered, and points to the important role of magnetohydrodynamic effects
in astrophysical systems. However, while it is known that QK flows are linearly
stable it remains unknown whether there exists a nonlinear transition to
turbulence, even for incompressible hydrodynamic systems. Some experiments
\cite{Ji_Nature, SchartmanAA, Edlund_PRE} and simulations
\cite{Monico} indicate that such a transition is not likely, while others
present evidence that suggests that a subcritical transition may exist
\cite{Paoletti} and some simulations find significant transient growth of
perturbations that may allow for nonlinear effects to enter \cite{Maretzke}.
Fluid experiments in other regimes of operation that are not astrophysically-
elevant have observed bi-stability \cite{Zimmerman,Huisman},
suggesting that should a similar mechanism exist for QK systems then a
subcritical pathway to turbulence may explain angular momentum transport in
accretion disks \cite{Lesur}. We show in this work that the influence of the
boundaries is intimately connected to the global structure of flows in Taylor-
ouette experiments and, by extension, is also related to the tendency of these
systems to generate and sustain turbulence.

One of the long-standing challenges of Taylor-Couette experiments in the
quest to understand angular momentum transport in astrophysically-relevant
flows has been the parasitic presence of Ekman circulation (secondary
circulation) induced by the mismatch between the fluid velocity and the solid
body rotation of the axial boundaries. A significant reduction in Ekman
circulation has been realized in experiment by using axial boundaries that are
split into multiple rings capable of differential rotation. Under particular
boundary conditions, azimuthal velocity profiles of the fluid can be generated
that very nearly match that of ideal Couette rotation \cite{Ji_Nature,
Edlund_PRE, Burin06, Schartman_RSI}, the rotation profile that is expected in
the absence of axial influences for a constant radial flux of angular momentum,
and has been observed to hold over a wide range of Reynolds numbers
\cite{Schartman_RSI}. In contrast, studies in the ``classical'' configuration
where the axial boundaries rotate with the outer cylinder have shown
performance that further deviates from ideal Couette as the Reynolds number
is increased \cite{Nordsiek}. Such trends are revealing of whether these
systems are dominated by boundary interactions or internal dynamics, a
distinction with important consequences for the applicability of such
experiments to interpretation of astrophysical systems, especially at large
Reynolds numbers. First, through the experiments reported here we identify
two necessary criteria that define constraints on the boundary configurations
that allow near-ideal flows to develop. We then discuss the competing roles of
radial (Stewartson) boundary layers and axial (Ekman) boundary layers, from
which we develop a model that describes the quantitative departure of the
rotation profiles from ideal Couette flow as a function of the angular
momentum fluxes through the boundaries.

\section{Experimental Apparatus}

\begin{figure}[t]
\begin{centering}
\includegraphics[width=10cm]{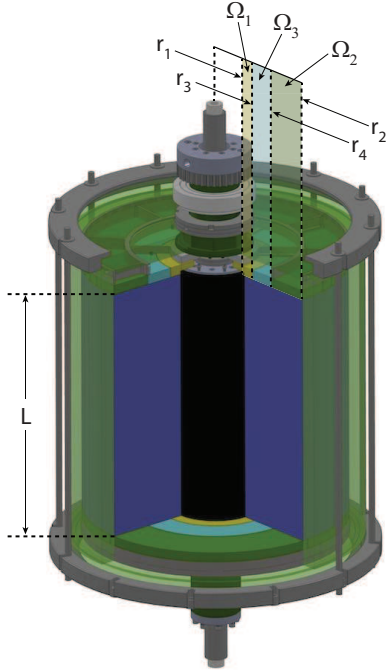}
\caption{\label{fig1} (color online) Illustration of the HTX device at PPPL
showing the segmented axial boundaries (yellow, blue and green segments),
the inner cylinder in black, the outer cylinder in green and the working fluid in
dark blue.}
\end{centering}
\end{figure}

A Taylor-Couette (TC) device is a system of coaxial cylinders that rotate
independently of each other with the experimental fluid region between. The
TC apparatus used in these studies, called the Hydrodynamic Turbulence
Experiment (HTX), is a modified version of the classical device in that the
axial boundaries in HTX are segmented to allow differential rotation across
the boundaries \cite{Edlund_PRE}. The inner cylinder radius is $r_1 = 6.9$ cm
and the outer cylinder radius is $r_2 = 20.3$ cm. The inner radius and outer
radius of the independent rings are defined by the parameters $r_3$ and
$r_4$, respectively. The axial length of the experimental volume is $L = 39.8$
cm, giving an aspect ratio of $\Gamma = L/(r_2-r_1) = 2.97$ (see
{Fig.}~\ref{fig1}). Corresponding components on the top and bottom are
driven by the same motor so that the system is up-down symmetric. The
angular velocities of the inner cylinder, outer cylinder and rings are identified
by $\Omega_1$, $\Omega_2$ and $\Omega_3$, respectively. Rotary
encoders report the speed of the motors to the control system.
 
A laser Doppler velocimeter (LDV) diagnostic system is used to measure the
local, azimuthal velocity ($v_\theta$), which in the experiments reported here
were measured at the midplane of the device. The LDV system is calibrated by
measuring fluid flow in solid body for which, after spin-up, the only velocity
component is $v_\theta$, which is a unique function of the motor speeds. For
these studies we define the global shear Reynolds number as $\text{Re}_s =
r_g^2 \Delta \Omega/\nu$, where $r_g = (r_1 r_2)^{1/2}$ is the geometric-
ean radius, $\Delta \Omega = \Omega_1 - \Omega_2$ and $\nu$ is the
kinematic viscosity, approximately $1\times 10^{-6}$ m$^2$/s for water.

\section{Observations}

\begin{figure}[t!]
\begin{centering}
\includegraphics[width=10cm]{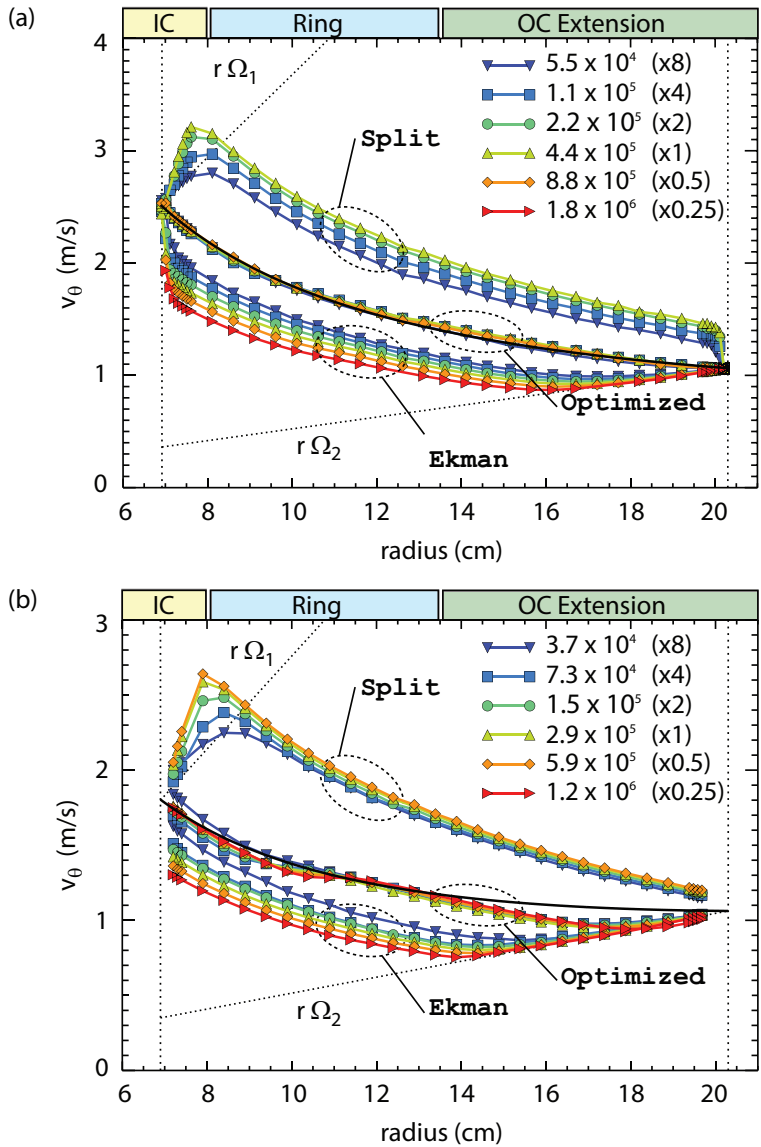}
\caption{\label{fig2} (color online) Scaled measurements of $v_\theta$ at
various $\text{Re}_s$ for (a) $q=1.8$ and (b) $q=1.5$.}
\end{centering}
\end{figure}

The core observation of this work is summarized in {Fig.}~\ref{fig2},
presenting flows under two different values of normalized shear ($q=1.8$ and
$q=1.5$) and three different boundary conditions: \texttt{Split},
\texttt{Optimized}, and \texttt{Ekman}. For the $q=1.8$ cases these
configuration speeds, reported as {$\Omega_1$-$\Omega_3$-$\Omega_2$},
are multiples of {350-350-50} RPM (\texttt{Split}), {350-185-50} RPM
(\texttt{Optimized}), and {350-50-50} RPM (\texttt{Ekman}). For the $q=1.5$
these are {250-250-50} RPM (\texttt{Split}), {250-140-50} RPM
(\texttt{Optimized}), and {250-50-50} RPM (\texttt{Ekman}). The reference
profile for these studies is described by ideal Couette rotation, defined in terms
of angular velocity as $\Omega_\text{C} (r)= \Omega_A + \Omega_B (r_g/r)^2$, where $\Omega_A = (r_2^2\Omega_2 -
r_1^2\Omega_1)/(r_2^2-r_1^2)$, $\Omega_B = r_g^2 (\Omega_1 -
\Omega_2)/(r_2^2-r_1^2)$. As a function of azimuthal velocity the Couette
solution is $v_\text{C}(r) = r \Omega_\text{C}$. It is interesting that while
the \texttt{Ekman} and \texttt{Split} configurations exhibit progressive
departure from ideal Couette as the Reynolds number is increased, the shape
of the \texttt{Optimized} cases is nearly invariant with respect to scaling of
the Reynolds number.

\subsection{Necessary conditions for near-ideal flows}

Recognizing that the axial boundaries in the \texttt{Split} cases will tend to
increase the angular momentum flux to the bulk, and conversely for the
\texttt{Ekman} cases, we begin by considering the ansatz that the balance of
angular momentum fluxes across the axial boundaries determines the
deviation from ideal Couette. A continuous variation of $\Omega_3$ from
$\Omega_2$ to $\Omega_1$ would then suggest that there is some
intermediate state for which the profile must pass close to ideal Couette
rotation. We now address whether there is a method for predicting this optimal
value of $\Omega_3$.

The transition between the bulk flow and the walls of the TC device, which
may move at very different speeds, occurs over thin boundary layers. The
structure of these boundary layers depends on whether we are considering the
balance of forces in the axial or radial directions. The bulk flow transitions to
the axial boundary speeds over Ekman boundary layers whose thickness scales
like $\delta_\text{E} \sim r \text{Re}_b^{-1/2}$, and in the ideal model of
radial boundaries, over Stewartson boundary layers that scale like
$\delta_\text{S} \sim r \text{Re}_b^{-1/4}$, where $\text{Re}_b = r^2
\Omega_b /\nu$ is a local Reynolds number particular to the boundary location
$r$ with boundary speed $\Omega_b$ \cite{BH73, Baker67, Stewartson}. The
fluxes of angular momentum across these boundary layers, being inversely
proportional to the boundary layer thickness, do not scale proportionally, and
hence we anticipate that self-similarity of the global properties need not be
preserved as the Reynolds number is scaled. Thus, while the observed
variation in the shape of the \texttt{Ekman} and \texttt{Split} profiles is
expected, it comes as some surprise that the \texttt{Optimized} profiles are
effectively independent of the Reynolds number.

The experiments reported here showed no significant temporal variation in the
mean values, hence, it can be stated that in steady-state the net flux of
angular momentum into these flows must sum to zero, that is, $\Phi_1 +
\Phi_2 + 2\Phi_z = 0$ where $\Phi_1$ and $\Phi_2$ are the fluxes integrated
over the inner and outer cylinders, respectively, and $\Phi_z$ is the axial flux
integrated over each axial boundary (with inward towards the fluid defined as
a positive flux). The ideal Couette profile is the response to a constant radial
flux of angular momentum, implying that $\Phi_2 = -\Phi_1$ and that the axial
fluxes of angular momentum are everywhere zero. Such a state can be
imagined in a TC system with free-slip conditions on the axial boundaries, or
with a continuously variable boundary that can perfectly match the ideal
Couette profile, conditions that will result in vanishing stress at the axial
boundaries. The radial flux of angular momentum under these conditions
(which we call $\Phi_\text{C}$, the Couette flux) has a magnitude equal to  

\begin{equation}
\Phi_\text{C} = \Phi_0 \frac{r_g^2}{r_2^2 - r_1^2} \text{Re}_s,
\label{phic}
\end{equation}

\noindent where $\Phi_0 = 4 \pi \rho \nu^2 L$. As an aside, it is interesting to
note that $\Phi_C = \Phi_0 \; \text{Re}_s$ when the radius ratio ($r_2/r_1$)
is equal to the golden ratio.

Even though there is, in general, a large mismatch between fluid and axial
boundary speeds, we find that the flows under the \texttt{Optimized}
boundary conditions very closely approximate the ideal Couette profile, and
therefore have a nearly constant radial flux of angular momentum. We now
show that rather than satisfying the condition that the axial flux of angular
momentum everywhere vanishes, these flows satisfy a much weaker
constraint: it is the surface integral of the axial angular momentum flux that
vanishes, that is, $\Phi_z \approx 0$.

To calculate the axial flux of angular momentum, being proportional to $d
\Omega/dz$, we make the assumption that $d \Omega/dz \approx \Delta
\Omega_E / \delta_E$, where $\Delta \Omega_E = \Omega_b-\Omega$ is the
difference of the boundary angular velocity ($\Omega_b$) and the fluid
angular velocity just inside the Ekman boundary layer ($\Omega$), and that
$\delta_E = \alpha_E \left( \nu/\Omega_b\right) ^{1/2}$ represents the
thickness of an Ekman boundary layer with an unknown numerical constant
$\alpha_E$.  Defining the normalized quantities $\omega=\Omega/\Delta
\Omega$ (recalling, $\Delta \Omega = \Omega_1 - \Omega_2$), $\omega_b
= \Omega_b/\Delta \Omega$, $s_1 = r_1/r_g$ and $s_2=r_2/r_g$, we have

\begin{equation}
\Phi_z = \Phi_0 \frac{r_g}{ 2\alpha_E L} \text{Re}_s^{3/2}
\int_{s_1}^{s_2} \left(\omega_b - \omega \right) \omega_{b}^{1/2} s^3 ds.
\label{phiz}
\end{equation} 

That the radial flux of angular momentum under ideal Couette rotation
[{Eq.}~\ref{phic}] scales like $\text{Re}_s$, whereas $\Phi_z$ scales like
$\text{Re}_s^{3/2}$ means that unless steps are taken to force the integral
in {Eq.}~\ref{phiz} to zero, there will always exist a Reynolds number beyond
which the axial flux will overwhelm the Couette flux and cause the flow to
depart from ideal rotation, regardless of aspect ratio. From {Eq.}~\ref{phiz}
we can immediately conclude that TC devices of the ``classical'' geometry
with the axial boundaries co-rotating with either the inner cylinder or outer
cylinder will always have non-zero $\Phi_z$ and can therefore never be a good
model for astrophysical systems at sufficiently high Reynolds numbers to be of
interest. In looking for the conditions under which ideal Couette flow may be
generated, we set $\omega=\omega_\text{C}$ and search for the boundary
conditions (the $\omega_b$) that cause the integral of {Eq.}~\ref{phiz} to
vanish. 

\begin{figure}[t]
\begin{centering}
\includegraphics[width=14cm]{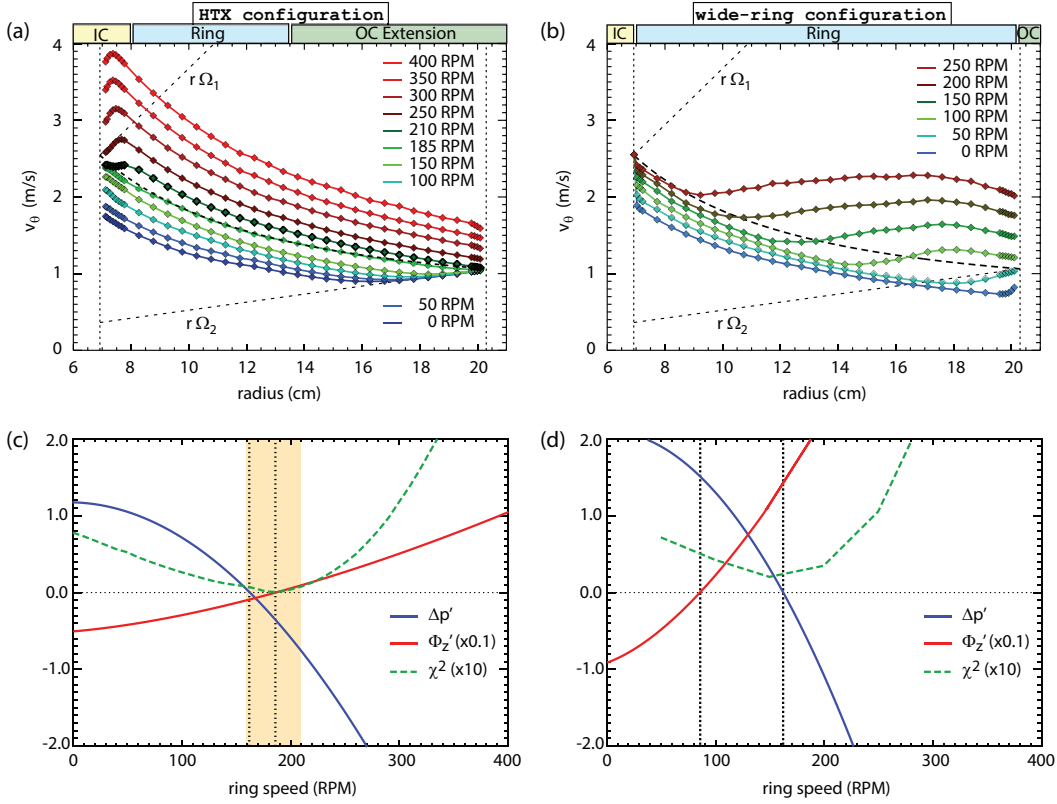}
\caption{\label{fig3}  (color online) Experiments conducted in $q=1.8$ flows
showing $v_\theta$ as a function of $\Omega_3$ for the (a) \texttt{HTX} and
(b) \texttt{wide-ring} geometries, progressing from low ring speeds (bottom
curves) to higher ring speeds (upper curves). In panels (c) and (d) the
experimental $\chi^2$ (dashed green), for the measured profiles relative to
the ideal Coutte profile for these boundary conditions, is plotted against the
calculated axial flux of angular momentum from {Eq.}~\ref{phiz} (red,
increasing function) and the pressure differential from {Eq.}~\ref{dp} (blue,
decreasing function) for the \texttt{HTX} and \texttt{wide-ring} geometries,
respectively. $\Phi_z$ is normalized by $\Phi_C$ and $\Delta p$ is normalized
by the kinetic energy density for ideal Couette flow. The vertical dotted lines
indicate the zeros of $\Delta p$ and $\Phi_z$.}
\end{centering}
\end{figure}

Figure \ref{fig3} summarizes fluid velocity measurements over a scan of
$\Omega_3$ for $q=1.8$ flows for two axial boundary configurations: the
\texttt{HTX} configuration with $r_3=7.8$ cm and $r_4=13.5$ cm, and a
\texttt{wide-ring} configuration with rings that span the entire radial gap.
These experiments show that the \texttt{HTX} configuration produces an
exceptional match to the ideal Couette profile for a narrow range of ring
speeds centered about $185$ RPM, with very low fluctuation levels spanning
the entire gap \cite{Edlund_PRE}. Importantly, the boundary conditions that
generate near-ideal flows for the \texttt{HTX} configuration include the zero
crossing of $\Phi_z$ [{Fig.} \ref{fig3}(c)]. However, the \texttt{wide-ring}
case also has a zero in $\Phi_z$, near $85$ RPM [{Fig.} \ref{fig3}(d)], yet its
flows never resemble ideal Couette, indicating that the vanishing of $\Phi_z$ is
a necessary but not sufficient condition for achieving near-ideal flows.

\begin{figure}[t]
\begin{centering}
\includegraphics[width=14cm]{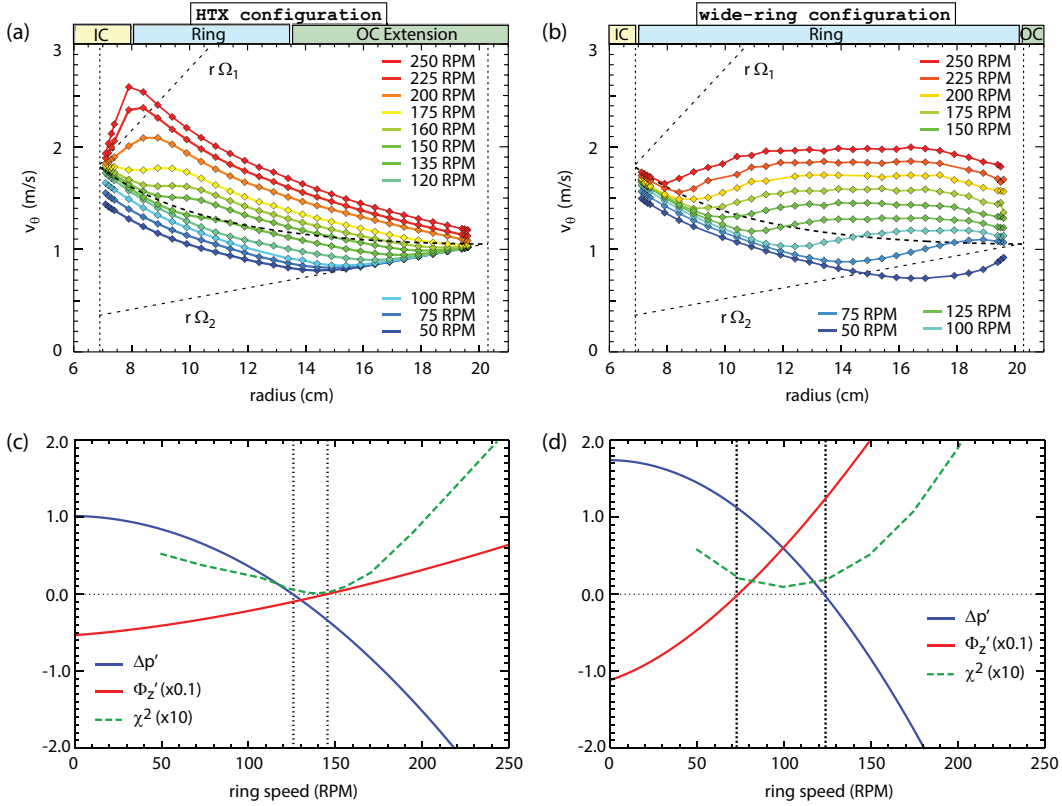}
\caption{\label{fig4}  (color online) Same as {Fig.}~\ref{fig3} except for
$q=1.5$.}
\end{centering}
\end{figure}

The role of the axial boundaries in dictating global performance has also been
interpreted through the competition of pressures from the bulk flow and from
the boundary flow that is viscously coupled to the axial boundaries
\cite{Obabko,Czarny}. The tendency to drive secondary flows was shown to be
consistent with whether the bulk pressure is larger than (Ekman circulation) or
smaller than (anti-Ekman circulation) the boundary pressure. Following the
intuition motivated by these simulations, we define a function $\Delta p$ that
characterizes the average pressure difference between ideal Couette rotation
and boundary rotation,
 
\begin{equation}
\Delta p =  \frac{\rho r_g^3 \Delta \Omega^2}{\Delta r} \int_{s_1}^{s_2}
\int_{s_1}^{s} \left(\omega_\text{C}^2 - \omega_b^2 \right) \: s' \: ds' \:
ds.
\label{dp}
\end{equation}
  
For the \texttt{wide-ring} configuration the zeros of $\Delta p$ and $\Phi_z$
are widely separated, meaning that no circumstance exists where pressure
balance and zero net axial flux can be simultaneously satisfied. In contrast,
the \texttt{HTX} configuration has zeros of these functions that are nearly
coincident and fall within the operating range in which small fluctuations
were observed in {Ref.}~\cite{Edlund_PRE}. Other boundary configurations
with $q=1.5$ also have nearly coincident zeros of $\Phi_z$ and $\Delta p$, in
remarkably good agreement with experiments; see {Fig.} \ref{fig4}. Note
that the requirement of coincident or nearly-coincident zeros represents an
effective third constraint. These studies suggest that the requirements of
having nearly coincident zeros of $\Phi_z$ and $\Delta p$ for the generation
of ideal flows may represent necessary and sufficient conditions. Additional
research conducted over a wider range of geometries and shear conditions
will provide a stronger test of this hypothesis.
   
\subsection{Departure of flows from ideal Couette}
    
While the coincident vanishing of $\Phi_z$ and $\Delta p$ define what may be
necessary and sufficient conditions for ideal flows to develop, they do not
reveal how QK systems should behave under non-optimized conditions. A
general model of the fluid response to forcing by the boundaries must account
for the fluxes across both axial and radial boundary layers. As there does not
exist a theory of Stewartson boundary layers under the conditions of QK
rotation at large Reynolds numbers, we cannot rely on results derived from a
linear analysis of perturbative differences in rotational speeds
\cite{Stewartson}, especially for experimental conditions where the
Stewartson boundary layers are turbulent \cite{Edlund_PRE}. Instead, we
assume a generalized scaling of the Stewartson boundary layer thickness of
the form $\delta_\text{S} = \alpha_\text{S} r (r/L) \text{Re}_b^\beta$,
taking the exponent of the Reynolds number ($\beta$) as a free-parameter to
be determined from the observed scalings in {Fig.}~\ref{fig2}. While the
numerical factors $\alpha_\text{E}$ and $\alpha_\text{S}$ are introduced
through the definitions of the boundary layer thicknesses we cannot measure
the boundary layers directly, and therefore we must interpret their meaning as
a measure of the effectiveness of angular momentum flux.

We begin this analysis by considering the \texttt{Ekman} configurations from
{Fig.}~\ref{fig2} where we note that $v_\theta$ transitions to solid body
rotation at $\Omega_2$ near the outer cylinder, implying that $\Phi_2 \approx
0$. With a jump in azimuthal velocity (a negative $\Delta v$) occurring over a
Stewartson layer at the inner cylinder, the appropriate form for $\Phi_1$ is

\begin{equation}
\Phi_1 = - \Phi_0 \frac{c_1}{2 \alpha_S}  \: \text{Re}_s^{\beta} \: \frac{L
\Delta v}{\nu},
\label{phi1}
\end{equation}

\noindent where $c_1 = (r_1 \Omega_1/r_2 \Delta \Omega)^\beta$ is a
result of converting from a representation of the boundary layer thickness that
depends on the local $\text{Re}_b$ to a global $\text{Re}_s$. We
approximate these flows with a piecewise function of the form $v_\theta(r_a <
r <  r_t) = v_\text{C} + \Delta v$ and $v_\theta(r_t < r <  r_2) = r
\Omega_2$, where $r_t$ is the transition radius, similar to the identification of
separate flow regions in the recent work of Nordsiek \textit{et al.}
\cite{Nordsiek}. Substituting this rotation profile into {Eq.} \ref{phiz} (using
$\omega = v_\theta/r\Delta\Omega$) and using {Eq.} \ref{phi1} for the
radial flux at $r_1$, we solve the global balance of angular momentum,
$\Phi_1 + 2 \Phi_z = 0$, for the dimensionless $\Delta v'_\text{neg}=\Delta
v/ r_g \Delta \Omega$ as a function of $\text{Re}_s$. Noting that $\Phi_z$
has two parts given this representation, our solution has three terms and is of
the form

\begin{equation}
\Delta v'_\text{neg} = \frac{f_1}{ \alpha \: c_1 \: \frac{L^2}{2 r_g^2}\:
\text{Re}_s^{\beta - 1/2} + f_2},
\label{dv1}
\end{equation}

\noindent where $\alpha = \alpha_\text{E}/\alpha_\text{S}$, and the integral
expressions $f_1$ and $f_2$ are defined as

\begin{equation}
f_1 = \int_{s_1}^{s_t}\left(\omega_b - \omega_\text{C}\right)
\omega_b^{1/2} s^3 ds
\label{f1}
\end{equation}

\begin{equation}
f_2 = \int_{s_1}^{s_t} \omega_b^{1/2} s^2 ds.
\label{f2}
\end{equation}

\begin{figure}[H]
\begin{centering}
\includegraphics[width=10cm]{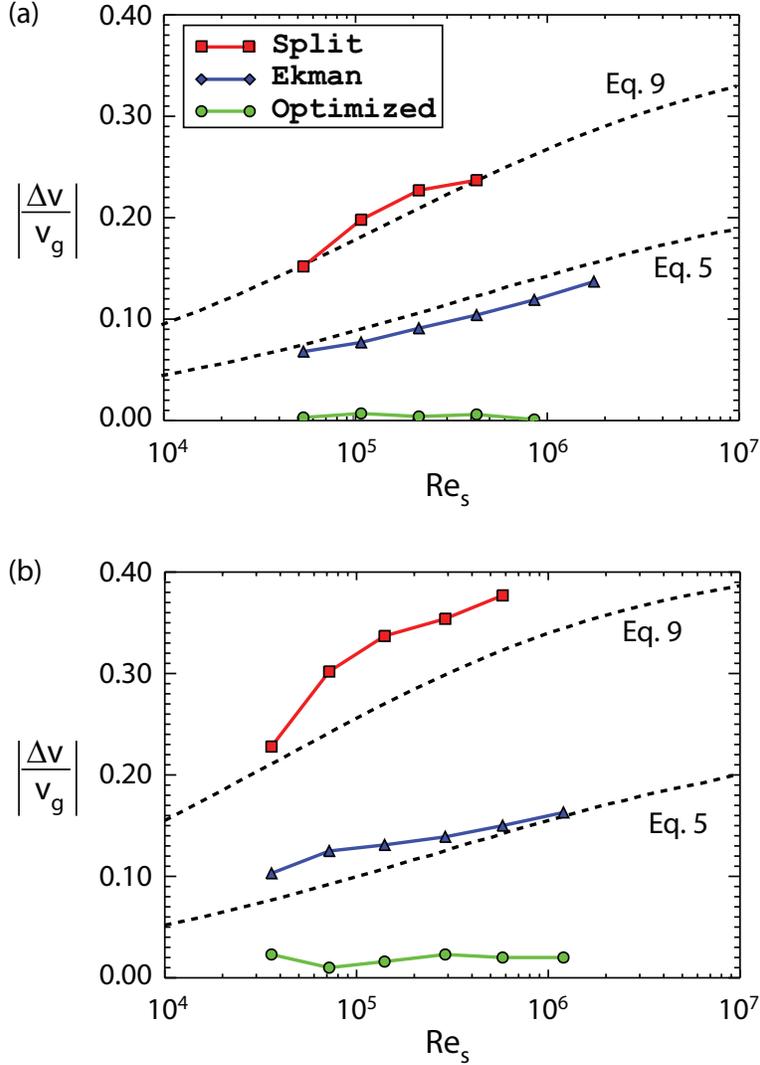}
\caption{\label{fig5} (color online) Scaling of the velocity differential $\Delta
v$ for the three groups of boundary conditions from {Fig.}~\ref{fig2} for (a)
$q=1.8$ and (b) $q=1.5$. The dashed lines in the figure are the solutions to
{Eqs.} \ref{dv1} and \ref{dv2} with $\beta = 0.08$ and
$\alpha_\text{E}/\alpha_\text{S} = 15$.}
\end{centering}
\end{figure}
 
\noindent Equation \ref{dv1} is a transcendental expression in $\Delta v$
since the limits of the integrals in $f_1$ and $f_2$ are functions of $r_t$
which itself depends on $\Delta v$. Equation \ref{dv1} has only two free
parameters: $\beta$ and $\alpha=\alpha_E/\alpha_S$. Comparison of {Eq.}
\ref{dv1} with experimental measurements of $\Delta v$ is presented in
{Fig.}~\ref{fig5} for the $q=1.8$ and $q=1.5$ cases, where values of $\beta
\approx 0.08$ and $\alpha_E/\alpha_S \approx 15$ provide the best fit to the
experimental measurements.
 
This analysis can be extended to the case of the \texttt{Split} configuration
with positive $\Delta v$ by accounting for the different boundary conditions
and a slightly more complex flow structure. The most important feature of the
positive $\Delta v$ cases is that there is a nearly linear decrease in $\Delta v$
across the gap, where $\Delta v$ at $r_2$ is defined to be equal to $\epsilon
\Delta v$. In the $q=1.8$ cases $\Delta v$ decreases by about $60\%$ across
the gap ($\epsilon=0.4$), and by $80\%$ for $q=1.5$ ($\epsilon=0.2$),
nearly independent of Reynolds number. The radial flux at $r_2$ in terms of
the $\epsilon$ reduced $\Delta v$ is

\begin{equation}
\Phi_2 = \Phi_0 \: \frac{c_2}{2 \alpha_S} \text{Re}_s^{\beta} \: \frac{ L \:
\epsilon \: \Delta v}{\nu},
\label{phi2}
\end{equation}

\noindent where $c_2 = (r_2 \Omega_2/r_1 \Delta \Omega)^\beta$. The
solution for positive, normalized $\Delta v$ is

\begin{equation}
\Delta v'_{pos} = \frac{f_3}{\epsilon \: \alpha \: c_2 \: \frac{L^2}{2 r_g^2}
\: \text{Re}_s^{\beta - 1/2} + f_4},
\label{dv2}
\end{equation}

\noindent where, similar to {Eq.} \ref{dv1}, we have 

\begin{equation}
f_3 = \int_{s_t}^{s_2}\left(\omega_b - \omega_\text{C}\right)
\omega_b^{1/2} s^3 ds
\label{f3}
\end{equation}
\begin{equation}
f_4 =  \int_{s_t}^{s_2} \left[ \frac{s_2 - \epsilon s_t - (1-\epsilon) s }{s_2-
_t} \right] \omega_b^{1/2} s^2 ds.
\label{f4}
\end{equation}

\noindent For the \texttt{Split} cases, the term $s_t$ defines the transition
point from increasing to decreasing $\Delta v$, and the terms in square
brackets in {Eq.}\ref{f4} derive from the variation of $\Delta v$ with radius
(note that if $\Delta v$ were constant across the gap then $\epsilon =1$ and
the term in brackets reduces to unity, recovering {Eq.}~\ref{f2}).
Equation~\ref{dv2} is compared against the measured $\Delta v'$ in
{Fig.}~\ref{fig5} and is also found also to be in good agreement with
experiment when using the same values of $\alpha_E/\alpha_S$ and $\beta$
derived from the analysis of the \texttt{Ekman} cases.

The success of this model in reproducing the general features of the
departures from ideal Couette is remarkable given that it represents a simple
accounting of the angular momentum fluxes across the boundaries, ignoring
completely the complex internal dynamics of these flows, especially at large
Reynolds number where a substantial fraction of the volume exhibits turbulent
fluctuations \cite{Edlund_PRE}. What disagreement exists between
experiment and theory should not overshadow the success of this model in
reproducing the general trend of the velocity deviations, the relative amplitude
of the positive and negative $\Delta v$ cases, and the change in the amplitude
of the positive $\Delta v$ cases between $q=1.8$ and $q=1.5$. While the
model presented here uses a single set of parameters for both values of $q$,
better fits to the data can be found if we let these parameters depend on $q$.
It is also possible the these parameters may depend on geometry and
Reynolds number in a way that is not captured by the power-law scaling used
here.

Studies of the measured fluctuation levels arising from the Stewartson
boundary layers in {Ref.}~\cite{Edlund_PRE} found that the transition to
turbulent Stewartson boundary layers was consistent with the Taylor model of
low Reynolds number centrifugal instability. So while cannot offer a precise
explanation for why $\beta$ should take a value close to $0.1$, we reiterate
that the inferred scaling applies to the angular momentum flux which is a
combination of variations in the boundary layer thickness and an effective
viscosity within these layers that is modified by turbulent fluctuations. It is
beyond the scope and ability of these experiments to identify the separate
scalings of the turbulent viscosity and the thickness of the Stewartson
boundary layers, though perhaps future experiments will be able to provide
greater insight into this problem.

\section{Conclusions}

We conclude with an outlook to future physical and numerical experiments.
The ratio of $\Phi_z$ and $\Phi_\text{C}$ explored earlier can be recast as a
constraint on the aspect ratio $L/\Delta r$ as a function of the Reynolds
number. The relative contribution of $\Phi_z$ can be made arbitrarily small by
increasing the axial size of the system, so that to make $\Phi_z/\Phi_\text{C}
\approx 10^{-2}$ for an \texttt{Ekman} configuration, for example, one would
need an aspect ratio of order $100$ at a Reynolds number of $10^6$. Outside
of using very large aspect ratios, one can employ mechanical advantages as
we have done in HTX, such as extensions of the inner and outer cylinder or
independent rings. Despite the desire to simplify the mechanical design of such
modified TC devices, further analysis shows that the only way to achieve
simultaneous zeros in $\Phi_z$ and $\Delta p$ is through a design with at
least one independent ring. Greater insight into the balance of forces in quasi-
eplerian flows could be explored in future experiments in modified Taylor
Couette devices like HTX. In particular, further exploration of the Reynolds
number dependence on these deviations, the dependence of the model
parameters on $q$, and the structure of the Ekman cells would tell us much
more about the influence of the boundaries on global behavior.

We have shown that nearly ideal flows exhibit profile shape invariance under
scaling of the Reynolds number, an effect we interpret through the dual
conditions of vanishing axial angular momentum flux and vanishing pressure
differential that are nearly simultaneously satisfied, offering predictive
capability for selecting optimized boundary conditions and in experimental
design. The strongest piece of evidence in support of this model is the
prediction of self-similarity of the profiles with respect to scaling of the
Reynolds number only for cases in which the axial flux of angular momentum
and the pressure differential vanish nearly coincidentally, a prediction in
excellent agreement with the observations presented in {Fig.}~\ref{fig2}. It
should be reiterated that while the experiments for \texttt{Optimized} flows
show very small departures from ideal Couette, and whose interpretation is
congruent with the coincident zeros in $\Delta p$ and $\Phi_z$, this does not
strictly prove that these criteria represent true sufficient conditions. An
interesting problem for future studies would be to measure how the
departures from ideal Couette depend on the ``distance'' between these
zeros.

It is interesting to note that in all cases presented here the local axial fluxes of
angular momentum may be quite large given the substantial difference in
speed between the bulk fluid and the axial boundaries, and that it is only on
summation over the axial boundaries that zero net axial flux is realized for the
\texttt{Optimized} cases. Recalling that multiple experiments
\cite{Edlund_PRE, Schartman_RSI,Burin06} and simulations
\cite{Obabko,Avila} have observed a nearly uniform axial structure through
the bulk of the fluid volume, the existence of the large axial fluxes naturally
raises the question: What allows the bulk flows to depart from the solid body
rotation forced by the boundaries? Intuition based on the Taylor-Proudman
theorem for Rayleigh-stable flows, that is $d v_\theta/dz \approx 0$, would
suggest that the bulk should tend to follow the boundary. However, it should
be recalled that the Ekman boundary layers do not need to satisfy the Taylor-
roudman theorem because the axial gradient of $v_\theta$ is balanced with a
viscous diffusion of vorticity over the scale of the boundary layer thickness. 
Thus, the nearly uniform axial structure suggests that the large fluxes of angular
momentum that must be present due to the existence of the the Ekman
boundary layers are redistributed locally, perhaps by small Ekman cells that do
not extend far into the bulk of the fluid or even in the boundary layers
themselves. As has been shown in prior studies, the free-shear layer that
develops at the interface between the differentially rotating rings on the axial
boundaries is expected to be centrifugally unstable and may help to enhance
the redistribution of angular momentum locally \cite{Spence}.

Another problem for future experiments, both physical and numerical, will be
to explore in greater detail how the very large, local fluxes of angular
momentum are redistributed near the axial boundaries with only minor effect
on the bulk flow in the \texttt{Optimized} configurations. Non-optimized
boundaries, like the \texttt{Split} and \texttt{Ekman} configurations, show
progressive departure from the ideal Couette flow as the Reynolds number is
increased, in agreement with the expectations of a dominant axial flux of
angular momentum. A good test for numerical experiments will be to
accurately model the global behavior of the mean flows in TC experiments by
properly accounting for the angular momentum fluxes from the boundaries. We
believe that the boundary layer scalings presented here may enable
simulations to bootstrap to larger effective Reynolds numbers by using
specified boundary flux models, thus bypassing the need for very fine grids to
resolve the boundary layer structure directly. With the combination of recent
simulations of QK flows that can attain Reynolds numbers of the order of
$10^5$ \cite{Monico,Shi}, though with axially-periodic boundary
conditions, a specified-flux boundary model may allow simulations to reach
something resembling experimental conditions of large Reynolds number TC
flows.

\section*{Acknowledgement} We thank J. Goodman for sharing his thoughts
on angular momentum transport and boundary layers in our studies, and E.
Schartman, E. Gilson, and P. Sloboda in helping to keep the experiments
running smoothly. This work was supported by the U.S. Department of Energy,
Fusion Energy Sciences under contract DE-AC02-09CH11466 through the
Center for Momentum Transport \& Flow Organization in Plasmas and
Magentofluids (CMTFO).

\end{document}